# Predicting Accurate Hot Spots in a More Than Ten-Thousand-Core GPU with a Million-Time Speedup over FEM Enabled by a Physics-based Learning Algorithm

Lin Jian[1], Yu Liu, Ming-Cheng Cheng*
Department of Electrical and Computer Engineering, Clarkson University, Potsdam
NY 13699-5720, USA
{jiangl2, yuliu, mcheng}@clarkson.edu

[1]Current affiliation: Department of Electronic and Computer Engineering,
The Hong Kong University of Science and Technology, Kowloon, Hong Kong
jiangwilbert72@gmail.com

*Abstract*—**The classical proper orthogonal decomposition (POD) with the Galerkin projection (GP) has been revised for chip-level thermal simulation of microprocessors with a large number of cores. An ensemble POD-GP methodology (EnPOD-GP) is introduced to significantly improve the training effectiveness and prediction accuracy by dividing a large number of heat sources into heat source blocks (HSBs) each of which may contains one or a very small number of heat sources. Although very accurate, efficient and robust to any power map, EnPOD-GP suffers from intensive training for microprocessors with an enormous number of cores. A local-domain EnPOD-GP model (LEnPOD-GP) is thus proposed to further minimize the training burden. LEnPOD-GP utilizes the concepts of local domain truncation and generic building blocks to reduce the massive training data. LEnPOD-GP has been demonstrated on thermal simulation of NVIDIA Tesla Volta™ GV100, a GPU with more than 13,000 cores including FP32, FP64, INT32, and Tensor Cores. Due to the domain truncation for LEnPOD-GP, the least square error (*LSE*) is degraded but is still as small as 1.6% over the entire space and below 1.4% in the device layer when using 4 modes per HSB. When only the maximum temperature of the entire GPU is of interest, LEnPOD-GP offers a computing speed 1.1 million times faster than the FEM with a maximum error near 1.2°C over the entire simulation time.**

## I. Introduction

Demands for high performance computing have drastically increased in recent years due to the needs for scientific and engineering computing and the explosion of machine learning, data science, and artificial intelligence [1]-[3]. The integration of a large number of cores in microprocessors enabling massive parallelism and the reduction in technology nodes enhancing the operation frequency have been a viable solution to continue improving the computing performance. Both approaches to satisfy the computing demands have inevitably increased power density in microprocessors [4], and thus led to temperature escalation and hot spot formation. Higher power dissipation degrades not only the computing performance but also the reliability of microprocessors and thus shortens their lifespan [5], [6]. To minimize the serious thermal issues, effective thermal management techniques are needed, which require efficient and accurate thermal simulation tools for microprocessors.

Among the conventional thermal simulation approaches, popular models based on thermal circuits [7]-[9] and the Green's function [10]-[12] are considerably more efficient than direct numerical simulation (DNS) that requires a large number of degrees of freedom (DoF). Assumptions are however needed in these efficient approaches at the cost of accuracy and/or spatial resolution. For example, the Green's function is a spatial impulse response of a unit point heat source, which is not able to account for effects of various boundary conditions (BCs) except the adiabatic BC that can be included using the method of image [13]. In addition, the Green's function is limited to 2D steady state simulations of a single layer with a heat flux dissipation on the substrate to account for effects of 3D thermal flow from the device layer to the substrate boundary.

Although approaches based on thermal circuits are more efficient than the Green's function, hot-spot temperatures may not be captured accurately unless the RC thermal elements are taken smaller than sizes of hot spots, which are as small as the grid size in DNSs. Use of small elements in thermal circuits, however, leads to a large-dimension matrix equation (i.e., a large DoF), which then becomes as time consuming as DNSs. It has been observed that thermal circuits usually offer an accurate prediction over a small interval in time or near steady state [13]-[19] due to inaccurate distributed heat transfer resulting from approximation of lumped elements. To correct the limit, HotSpot [8] includes one scaling factor for all lumped thermal elements in the entire domain to adjust the time scale.

Taking an unconventional path, proper orthogonal decomposition (POD) [20], [21] of solution data, together with the Galerkin projection (GP) of the heat transfer equation, has recently been applied successfully to thermal simulations of microprocessors to achieve high efficiency and accuracy with

This work was supported by the National Science Foundation under Grant Nos: ECCS-2003307 and OAC- 2118079.

fine resolution [22], [23]. Using POD-GP, the thermal problem is projected from a physical domain onto a POD space represented by a finite set of optimal basis functions (or POD modes) trained by temperature data obtained from DNSs of the problem. The GP of the heat transfer equation onto the POD space is further applied to close the model, which incorporates heat transfer principles into the model. The training has previously been performed globally for POD-GP (hereafter named GPOD-GP) in the entire processor subjected to variation of dynamic power maps (PMs) to generate a set of global POD modes, where the dynamic PM provides the strengths and locations of dynamic heat sources in the processor induced by real-time workload. This classical POD-GP approach offers an accurate prediction of the dynamic thermal profile in a multi-core processor using a small number of modes if the dynamic power map (PM) is within the training range [22], [23]. In situations slightly outside the bounds of the training settings, the accuracy deteriorates; however, good accuracy can still be reached with more modes included [22]. This effective learning ability of GPOD-GP stems from the GP of the heat transfer equation, which enforces the physical (heat transfer) principles as a guidance to reach a good prediction. This is very different from the mainstream machine learning methods based on neural networks, whose predictions usually fail for situations beyond the training [24].

To generate effective POD modes for GPOD-GP to achieve efficiency and accuracy, dynamic PMs applied in the training must cover enough spatial variations in dynamic power source locations. This can be easily achieved for processors with a small number of cores. For processors with hundreds of cores, the intensive training effort become prohibitive, and an ensemble POD-GP model (EnPOD-GP) is proposed, which drastically simplifies the training to generates good quality data and enhances the accuracy of the model. For processors with considerably more cores, domain truncation is applied to data collection to develop a local EnPOD-GP model (LEnPOD-GP), together with generic building blocks, to further minimize the training effort.

## II. Classical Proper Orthogonal Decomposition with Galerkin Projection

Spatiotemporal temperature $T(\vec{r},t)$ can be represented by a linear combination of a set of basis functions $\eta_i$ (or modes),

$$T(\vec{r},t) = \sum_{i=1}^{M} a_i(t)\eta_i(\vec{r}), \quad (1)$$

where $a_i(t)$ is the weighting coefficient of $\eta(\vec{r})$ and $M$ is the selected number of modes (i.e., the DoF). The modes in the classical POD-GP reduced order model (or GPOD-GP) are trained globally by temperature data collected from the entire domain in dynamic DNSs subjected to parametric variations. By maximizing the mean square of the temperature data projection onto the POD modes over the entire simulation domain, the maximization process leads to an eigenvalue problem,

$$\int_{\Omega} \langle T(\vec{r},t) \otimes T(\vec{r}',t)\rangle\, \eta(\vec{r}')d\Omega = \lambda\eta(\vec{r}), \quad (2)$$

where $\eta(\vec{r})$ is the eigenfunction, $\lambda$ is the eigenvalue, $\otimes$ is the tensor operator and $\langle\cdot\rangle$ denotes the average over the training data sets.

To determine $a_i(t)$, one can perform the GP of the heat transfer equation onto each POD mode,

$$\int_{\Omega} \left(\eta_i \frac{\partial \rho C T}{\partial t} + \nabla\eta_i \cdot k\nabla T\right) d\Omega = \int_{\Omega} \eta_i P_d d\Omega + \int_{S} \eta_i k\nabla T \cdot d\vec{S}, \quad (3)$$

where $\rho$, $k$ and $C$ are the density, thermal conductivity and specific heat, respectively, and $P_d(\vec{r},t)$ is the interior power density and $d\vec{S}$ is the outward differential surface vector. A set of $M$-dimensional ordinary differential equations (ODEs) for $a_i(t)$ can be shown as,

$$\sum_{i=1}^{M} c_{i,j}\frac{da_i}{dt} + \sum_{i=1}^{M} g_{i,j}a_i = p_j, \qquad j = 1\ to\ M, \quad (4)$$

where $c_{i,j}$, $g_{i,j}$ and $p_j$ are elements of the thermal capacitance matrix, the thermal conductance matrix and the power vector in the POD space. These elements are defined in terms of integrals of $\eta$ and $\nabla\eta$, and their expressions are given in [23]. $p_j(t)$ in (4) accounts for the interior power dissipation and boundary heat flux in the POD space defined on the right hand side of (3). These elements in the POD space can be pre-evaluated from POD modes and save in a library. $T(\vec{r},t)$ in the simulation domain can be determined via (1) once $a_i(t)$ are solved from (4) with a selected DoF of $M$.

## III. Ensemble POD-GP model

### A. *EnPOD-GP Background*

To minimize the intensive computing effort needed in POD mode training for GPOD-GP, the number of heat sources in a training domain needs to be small enough. For processors with many cores, one can divide the heat sources (provided by FUs and cores) into several heat source blocks (HSBs), where each block consists of a small number of heat sources (e.g., 1 to 4). An individual POD-GP model (IPOD-GP) can be built for each HSB; i.e., POD modes for each IPOD-GP are trained separately from others for the entire processor, responding to random dynamic power excitations in each source block. Thus, each IPOD-GP offers the temperature solution induced by the power dissipated by the corresponding HSB. An ensemble POD-GP model (EnPOD-GP) can then be constructed by summing temperatures resulting from all IPOD-GPs in the entire processor using the superposition principle. $T(\vec{r},t)$ in the processor is then given as,

$$T(\vec{r},t) = \sum_{n=1}^{N_{hb}} \sum_{i=1}^{M_n} a_{n,i}(t)\eta_{n,i}(\vec{r}), \tag{5}$$

where the indices represent the $i$th mode of the $n$th HSB, $N_b$ is the total number of HSBs in the processor and the weighting coefficients for the the $n$th HSB can be expressed in a vector form as $\vec{a}_n = [a_{n,1}, a_{n,2}, a_{n,3}, \dots, a_{n,i}, \dots, a_{n,M_n}]^T$. In this study, $M_n = M$; i.e., the number of ODEs for each HSB is identical. The equivalent $i$th-mode eigenvalue for the entire processor is defined as,

$$\lambda_i^{eq} = \sum_{n=1}^{N_{hb}} \lambda_{n,i} w_n, \tag{6}$$

where $w_n$ is the area fraction of the $n$th HSB. Using EnPOD-GP, $a_{n,i}$ in (5) is solved from the $n$th set of ODEs in (4) induced by $n$th HSB. Although $N_{hb}$ sets of ODEs needed to be solved when using EnPOD-GP, each set is independent, i.e., the system matrix of the ODEs is highly sparse unlike GPOD-GP where none of the elements in the ODE system matrix is zero.

Eigenvalues represent the mean squared temperature information captured by $\eta$. Since $M_n = M$ in this study, the relative least square error ($LSE$) of the solution predicted by EnPOD-GP can be theoretically estimated as

$$LSE_{theo} = \sqrt{\sum_{i=M+1}^{N_s} \lambda_i^{eq} \Big/ \sum_{i=1}^{N_s} \lambda_i^{eq}}, \tag{7}$$

where $N_s$ is the number of temperature data sets for generating POD modes and there are same number of data samples in each HSBs in this study. Numerically, the $LSE$ induced by EnPOD-GP with respect to DNS can be estimated from the predicted temperature as

$$LSE = \sqrt{\sum_{i=1}^{N_t} \int_\Omega e^2(\vec{r},t_i) d\Omega \Big/ \sum_{i=1}^{N_t} \int_\Omega [T(\vec{r},t_i) - T_o]^2 \, d\Omega}, \tag{8}$$

where $e(\vec{r},t_i)$ is the temperature difference between the DNS and EnPOD-GP at the $i$th time step, $N_t$ is the total number of time steps, and $T_o$ is the ambient temperature.

Note that the number of heat sources in an HSB needs to be small enough to generate a set of robust POD modes. One can have the following choices: (i) each HSB representing one heat source induced by only one FU, (ii) each HSB consisting of a few FUs each having an individual power source, or (iii) each HSB represented by one heat source with a total power of several FUs. Among these 3 choices, the first two offer more localized power excitations to capture more realistic hot spots. The first choice leads to more sets of ODEs while each set includes a small number of modes. The second one requires fewer sets of ODEs; however, each set needs more modes to reach a similar accuracy. The last one offers more efficient training and fewer sets of ODEs but may estimate unrealistically low hot-spot temperatures. To optimize both accuracy and efficiency, combinations of these 3 choices can be applied over the entire chip, depending on the power levels of the heat sources and their sizes. To simplify the demonstration, only one heat source is used in each HSB in this investigation.

It should be pointed out that post processing via (5) in general takes considerably more time than solving ODEs in (4). Unlike any DNSs or thermal circuits that need to perform simulation over the entire simulation time and domain to be able obtain dynamic thermal solution, for POD-GP based approaches $T(\vec{r},t)$ in (5) can be selectively determined at any point in time and space. Most thermally related applications in microprocessors only need thermal information near high temperature regions, such as the device layer or the high-power FUs (e.g., cores). Some applications (e.g., thermal-aware tasking scheduling and reliability assessment) may only need the peak temperature distribution in high-power FUs or even just the peak temperature of the entire processors. In these situations, the EnPOD-GP computing speed could be improved by one or 2 orders of magnitude.

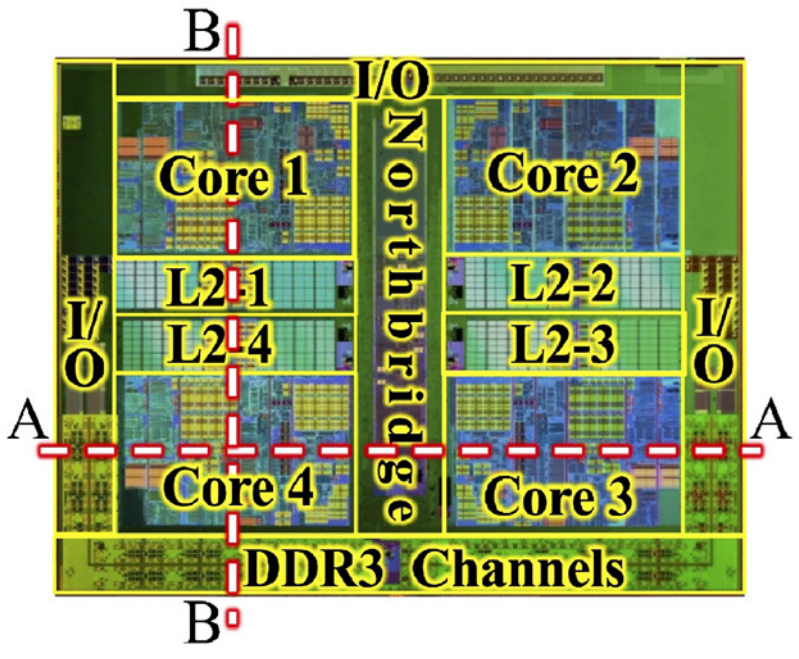

Fig. 1. Floorplan of the quad-core CPU, AMD ATHLON II X4 610e [25], with an area of $14 \times 12\ mm^2$.

### B. *Demonstration of EnPOD-GP in a Quad-Core CPU*

EnPOD-GP is demonstrated in dynamic thermal simulation of a quad-core CPU, AMD ATHLON II X4 610e CPU with its floorplan given in Fig. 1 [25]. To train the POD modes, temperature data are collected from the FEM simulation tool, FEniCS [26] with resolution of $0.093 \times 0.08 \times 0.042\ mm^3$ (a mesh of $150 \times 150 \times 17$). There are 13 FUs in this CPU, including four Cores, four L2 Caches, one Northbridge, one DDR3 Channels and three I/Os. PMs in this demonstration are generated from gem5 [27] and McPAT [28] using several benchmarks [29], where one uniform dynamic power source is generated in each FU. Due to the limit of gem5 that does not generate power in three I/O's and one DDR3 Channels, there are thus 9 FUs with power dissipation implemented in EnPOD-GP as HSBs. The EnPOD-GP system thus includes 9 sets of ODEs (thus 9 sets of POD modes with each set for an IPOD-GP ), and the number of ODEs (i.e., the number of modes) in each set can be determined by the eigenvalue spectrum generated from thermal data, based on the desired accuracy estimated from (7), as described below.

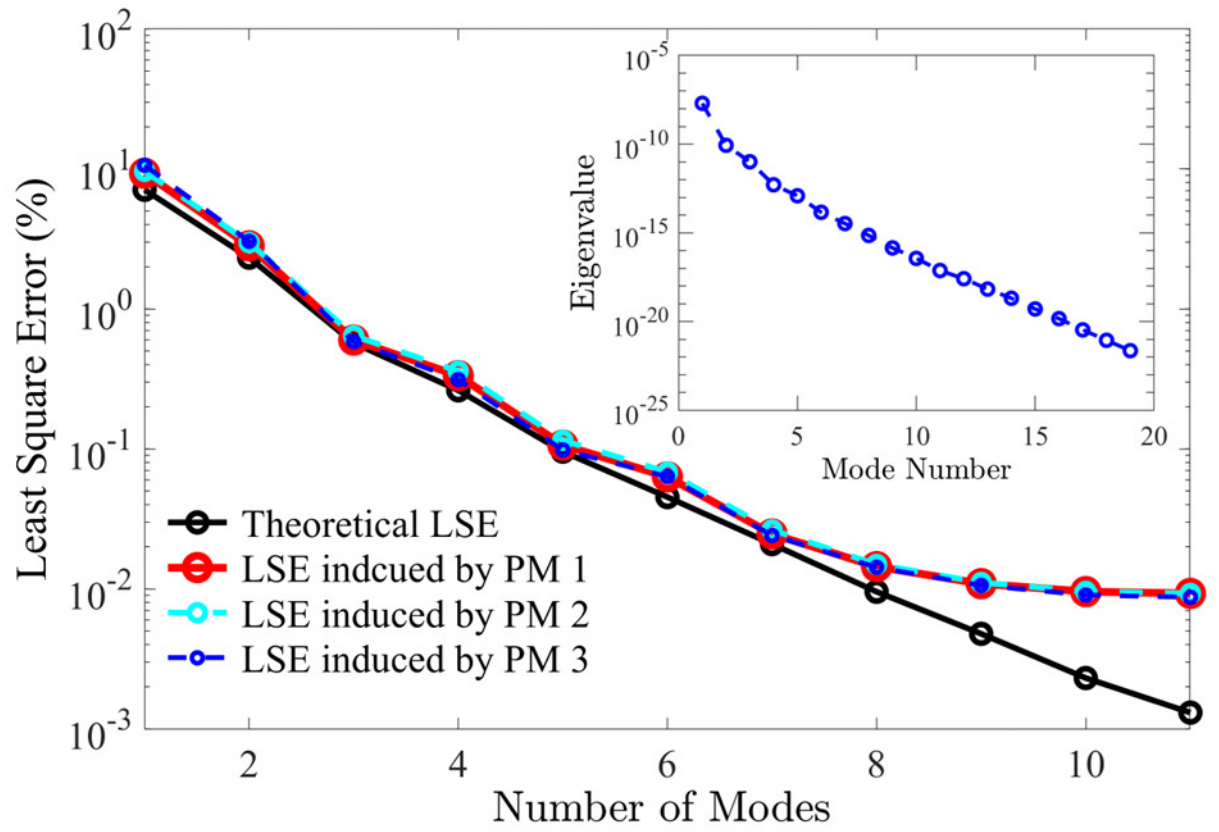

Fig. 2. *LSE* of EnPOD-GP vs. the number of modes per IPOD-GP for thermal simulation of the quad-core CPU. The inset includes the equivalent eigenvalue spectrum estimated in (6) using the eigenvalues for all 9 IPOD-GPs.

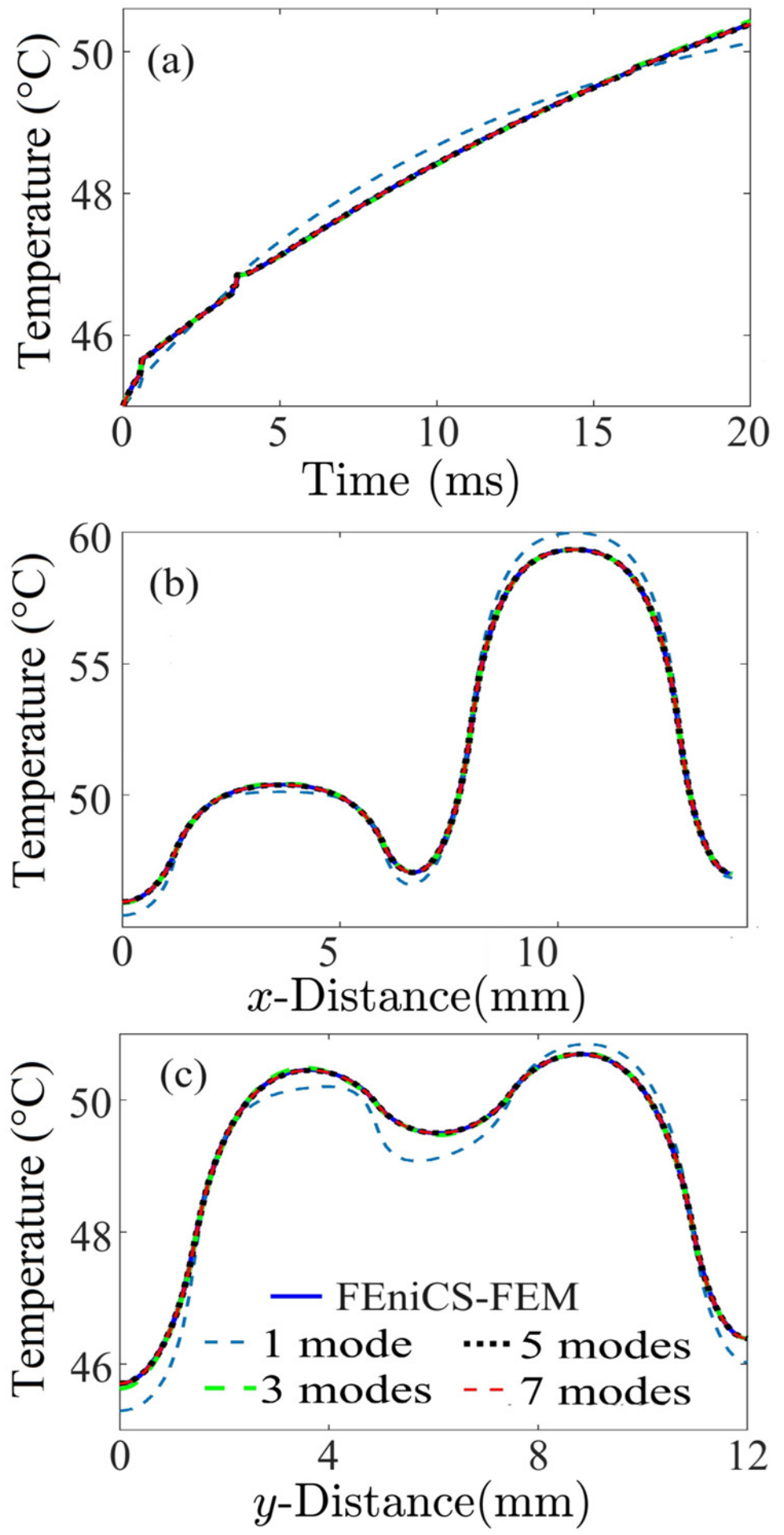

Fig. 3. (a) Dynamic temperature at the intersection of Paths A and B shown in Fig. 1. (b) and (c) Temperature distributions in $x$ along Path A and in $y$ along Path B. EnPOD-GP results are presented as the number of modes per IPOS-GP.

Using IPOD-GP to construct EnPOD-GP, each IPOD-GP is independent of each other. Each of the 9 sets of POD modes is thus trained separately by random dynamic power excitation. The equivalent eigenvalue of EnPOD-GP given in (6) for the processor is displayed in the inset of Fig. 2. The rapid eigenvalue reduction for EnPOD-GP observed in Fig. 2 thus leads to an $LSE_{theo}$ near 2.2%, 0.58% and 0.1% with just 2, 3 and 5 modes per HSB, respectively. Moreover, the $LSE$ evaluated from (8) in EnPOD-GP thermal simulations of the processor induced by 3 different PMs generated from gem5 [27], McPAT [28] and benchmarks [29] are nearly identical and accurately predicted by $LSE_{theo}$ until computer precision is reached. The $LSE$ stays near 0.01% beyond 8 modes. As illustrated in Fig. 3, excellent accuracy of EnPOD-GP with 3 modes per HSB (27 modes in total) is observed for thermal solution in time and space subjected to one of the PMs used in Fig. 2, compared to FEniCS-FEM simulation.

In addition to simple training of each individual HSB, results illustrated in Figs. 2 and 3 from EnPOD-GP simulation of the AMD quad-core CPU have demonstrated several advantages of EnPOD-GP. The training of POD modes with simple random power excitations is remarkably effective and leads to a robust ENPOD-GP methodology that is independent of dynamic and spatial power source variations. Using any PM, a very accurate prediction of spatiotemporal thermal solution can be achieved with just 2 or 3 modes per HSB (per FU in this case) and yet its $LSE$ can be pre-estimated accurately from (7).

The DoF needed for EnPOD-GP is the selected number of modes $M$ per HSB (or per IPOD-GP) multiplied by $N_{hb}$ (the number of HSBs). Thus, the DoF equals 27 in EnPOD-GP if 3 modes per IPOD-GP for 9 sets of ODEs are used. Compared to FEniCS-FEM, the reduction of DoF is 4 orders of magnitude ($150 \times 150 \times 17/27$). The decrease in computing time, compared to FEniCS-FEM simulation is near 2,600 times if 3 modes per HSB is used ($LSE \approx 0.78\%$) to evaluate the temperature in the entire process and 3,500 times if 2 modes per HSB are used ($LSE \approx 2.8\%$). In cases where only temperature at several points in space/time need to be evaluated from (5), at least a one-order reduction in computing speed can be achieved.

## IV. Local Ensemble POD-GP Model

Although the training is considerably simpler and more effective than GPOD-GP, the intensive training effort needed for EnPOD-GP still becomes intolerable when too many HSBs need to be trained in a microprocessor with thousand or more cores to generate numerous sets of POD modes. With some modifications described below based on concepts of thermal length $\lambda_{th}$ and generic building blocks, a local EnPOD-GP model (LEnPOD-GP) is proposed to significantly minimize the training effort for GPUs with thousand or more cores.

In order to describe the developed LEnPOD-GP model more clearly, a workflow diagram is included in Fig. 4, which illustrates each step needed to develop LEnPOD-GP, perform simulation in POD space, and then post process the solution in POD space to obtain temporospatial temperature in a GPU. This diagram is used throughout in this section to offer a better understanding of LEnPOD-GP.

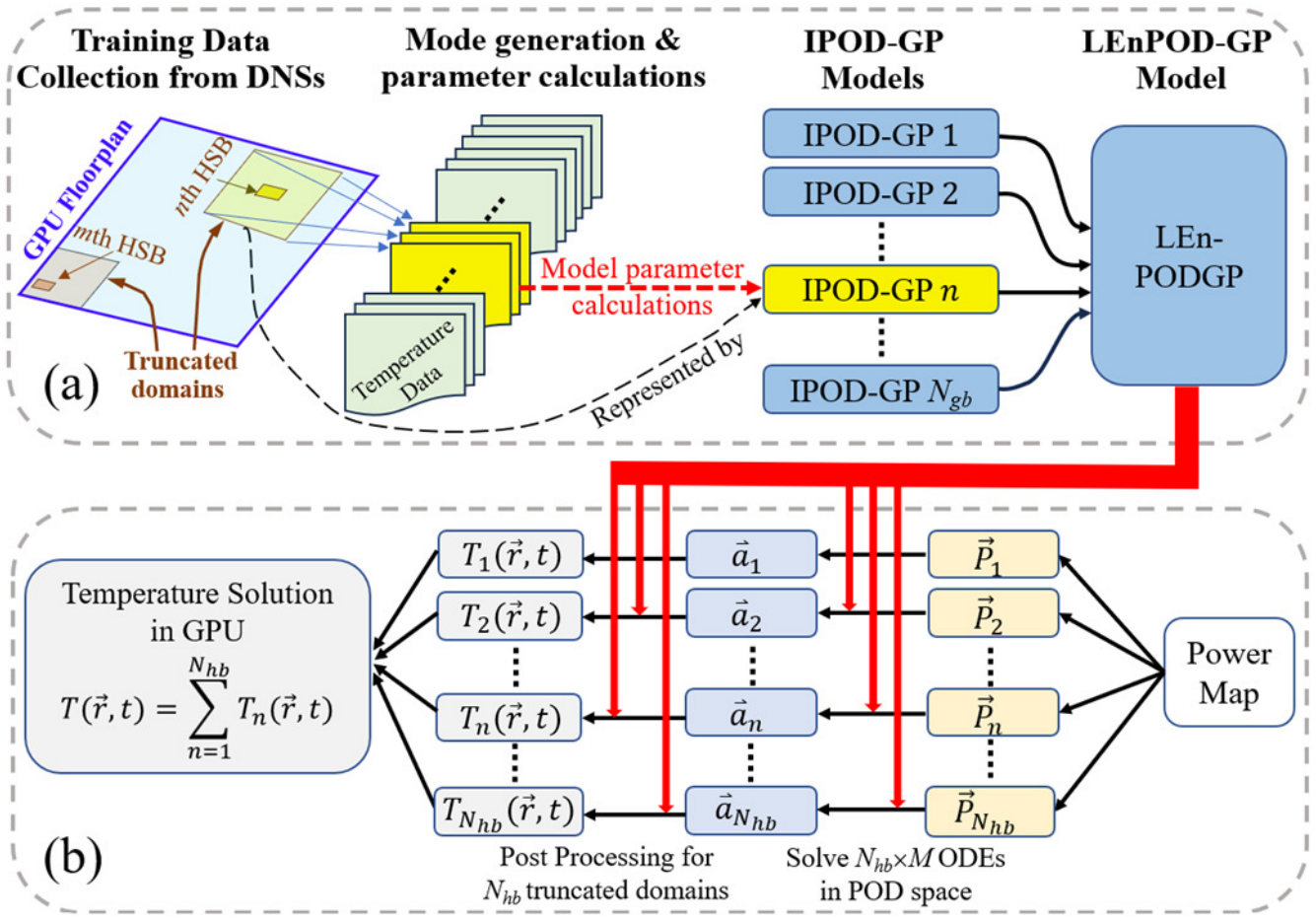

Fig. 4. Workflow chart for (a) development of LEnPOD-GP and (b) simulation (solving ODEs) in POD space and post processing to obtain $T(\vec{r},t)$ in a GPU. $N_{gb}$ is the number of generic building blocks each represented by an IPOD-GP model. However, $N_{hb}$ is the number of HSBs or truncated domains that are all covered by $N_{gb}$ IPOD-GPs. In this study for the GPU, $N_{gb}$=16 and $N_{hb}$ = 404. Also, $\vec{a}_n = [a_{n,1}, a_{n,2}, \ldots, a_{n,M}]^T$ and $\vec{P}_n = [p_{n,1}, p_{n,2}, \ldots, p_{n,M}]^T$ with $M$ as the number of modes used in each truncated domain to represent the solution.

## A. *LEnPOD-GP Background*

**Thermal Length:** Temperature induced by a heat source diffuses and decays in space. To simplify the characterization of the decreasing profile, thermal length $\lambda_{th}$ based on the concept of the exponential diffusion profile is used even though the profile is not exactly exponential. That is, $\lambda_{th}$ is defined as the distance measured vertically from an HSB boundary to a location where the temperature decreases to 36.8% of the HSB boundary temperature. Temperature induced by each HSB at a distance several thermal lengths away from the HSB can be neglected. A truncated local domain containing the HSB is defined for collecting the training data with the domain boundaries several thermal lengths away from the HSB unless the HSB is very close to GPU boundaries, as shown in the floorplan of Fig. 4(a) for 2 truncated domains. For example, the truncated local domain for the *m*th HSB is smaller because it is very close to the GPU boundaries. Using the training data collected from each truncated local domain, instead of the entire processor, one set of POD modes for each IPOD-GP (i.e., each HSB) is trained for the truncated domain, as indicated in Fig. 4(a). This substantially reduces the training effort for a large processor with a large number of cores. The distance between the HSB and the truncated local domain boundary can be varied to obtain the desired accuracy. In this investigation, five thermal lengths are taken to ensure the temperature induced by the HSB beyond the truncated domain is negligible.

**Generic building blocks:** Microprocessors in general consist of many repeated units or generic building blocks. For a processor with hundreds or thousands of cores/FUs, one IPOD-GP model for one generic block can then be trained to represent many identical truncated local domains containing identical HSBs, where each HSB may include several cores and/or FUs. This will significantly reduce the number of truncated training domains (i.e., $N_{gb}$,the number of generic building blocks or IPOD-GPs), which significantly minimizes the training effort and memory space. However, for some identical HSBs whose distance from any edge of the processor is less than the selected number of thermal lengths, separate training is needed to include the boundary effects on the processor edge for these truncated domains, such as the *m*th HSB in Fig. 4(a).

## B. *Training of LEnPOD-GP for Tesla V100 Volta GV 100 GPU*

The Tesla Volta™ GV100 GPU whose floorplan shown in Fig. 5 is selected to demonstrate the learning capability and accuracy for LEnPOD-GP. The Tesla GV100 GPU's thermal design power (TDP) is as high as 300W with a die size of 815 $mm^2$ and 21.1 billion transistors. There are 80 stream multiprocessor (SMs) in the GPU and each SM comprises four texture units and four processing blocks (PBs), where each PB consists of 16 FP32 Cores, 8 FP64 Cores, 16 INT32 Cores and two Tensor Cores. There are thus 13,440 cores in total. For this demonstration, 404 HSBs are selected, where each HSB represents each of 320 PBs, all 4 texture units within each of 80 SMs, one L2 cache, one high-speed hub or 2 memory interfaces. The areas of most cores/FUs in the selected GPU are considerably smaller than those in the quad-core CPU shown in Fig. 1. For example, there are 320 identical PBs (each includes 42 cores), and area of each PB is 18 times smaller than that of each of the CPU cores in Fig. 1 ($0.5 \times 1.79$ $mm^2$ vs. $4.78 \times 3.45$ $mm^2$). The chip area of Volta GV100 is however around 4.9 times larger than that of the AMD quad-core CPU ($28.6 \times 28.5$ $mm^2$ vs. $14 \times 12$ $mm^2$). This induces smaller-size hot spots in a considerably larger GPU than the CPU. To capture these smaller-size hot spots accurately in the training data collected from DNS, a finer mesh of $675 \times 673 \times 17$ is used in the GPU (compared to $150 \times 150 \times 17$ in the CPU).

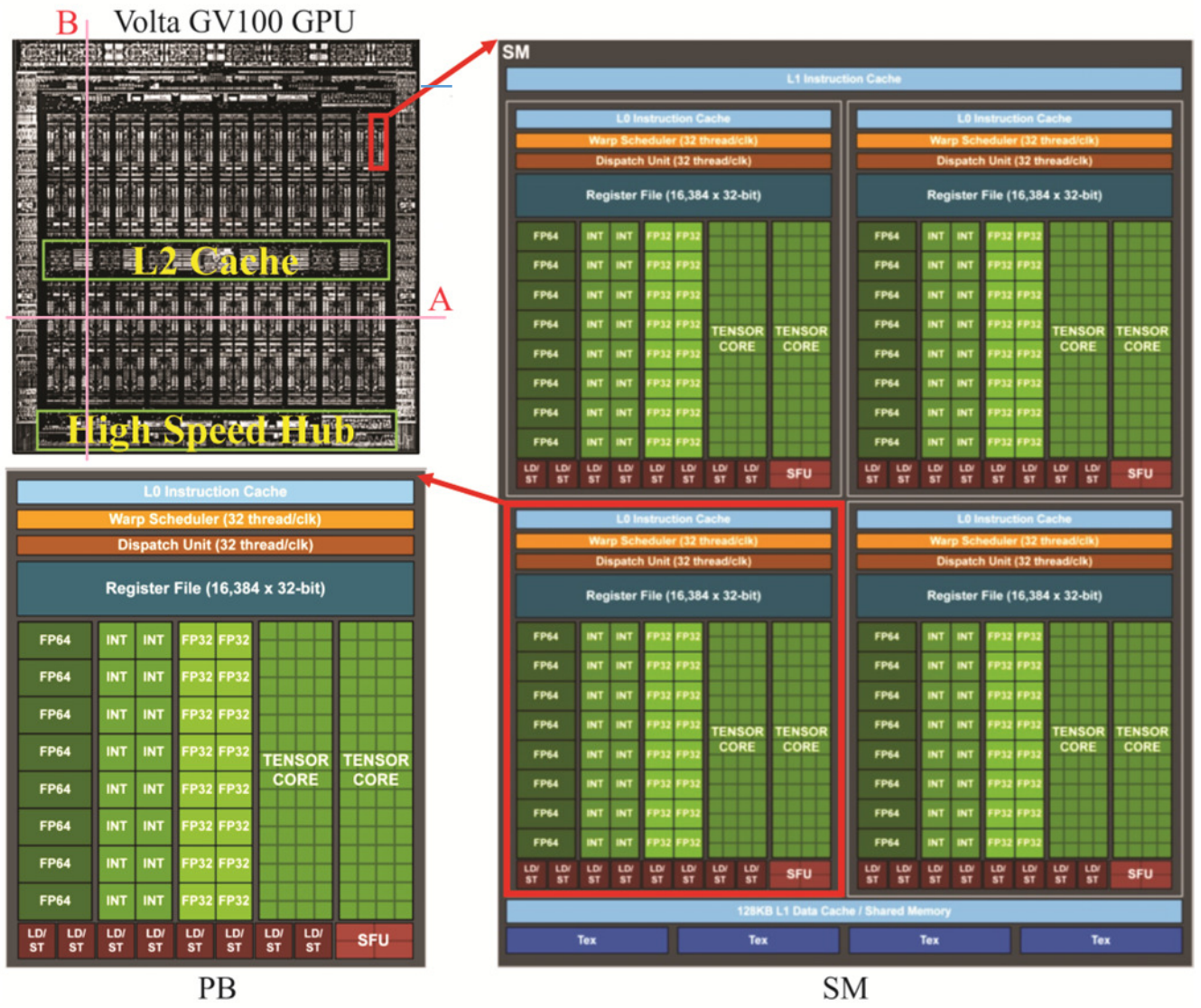

Fig. 5. floorplan of Tesla Volta GV100 GPU, together with zoom-in views of SM and PB [30]. The size of chip is $28.65 \times 28.5 \times 0.72$ $mm^3$. Lines A and B on the floorplan indicate the plotting paths for temperature profiles shown in Fig. 8.

As discussed above, thermal data are collected from the truncated domain for each of HSBs within $5\lambda_{th}$ beyond the HSB. $\lambda_{th}$ of an HSB on each side is influenced by chip thickness, materials, the HSB width vertical to the heat diffusion direction, and the aspect ratio of HSB. $\lambda_{th}$ is pre-evaluated in DNSs for widths between 1mm and 19 mm, and it is found that $\lambda_{th}$ varies from 0.8mm to 1.5mm for a chip thickness of 720 µm. The evaluated thermal length for each HSB is thus applied in data collection from the truncated generic local domain. Also, for some HSBs with a distance less than $5\lambda_{th}$ from any of the processor edges, such as the *m*th HSB in Fig. 4(a), the training is performed separately.

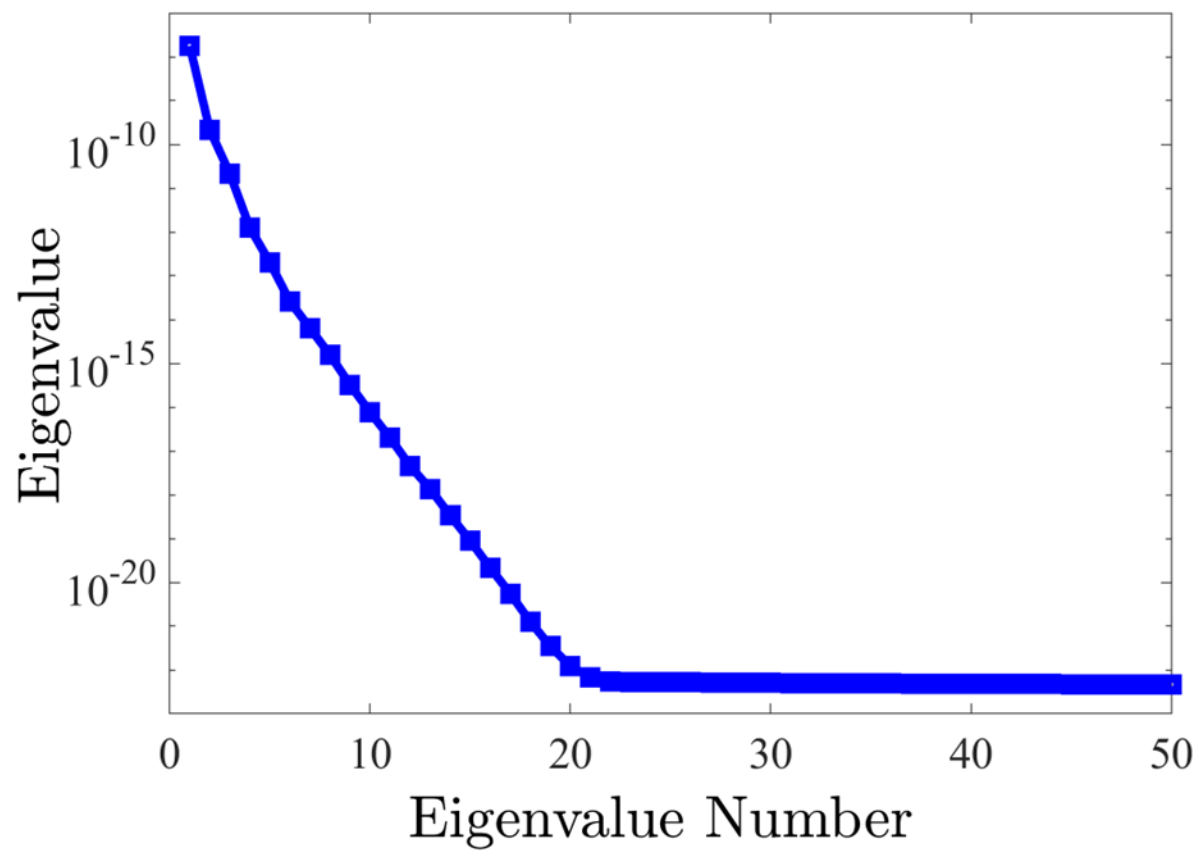

Fig. 6. Equivalent eigenvalues of LEnPOD-GP for thermal data collected from Tesla Volta GV100 GPU.

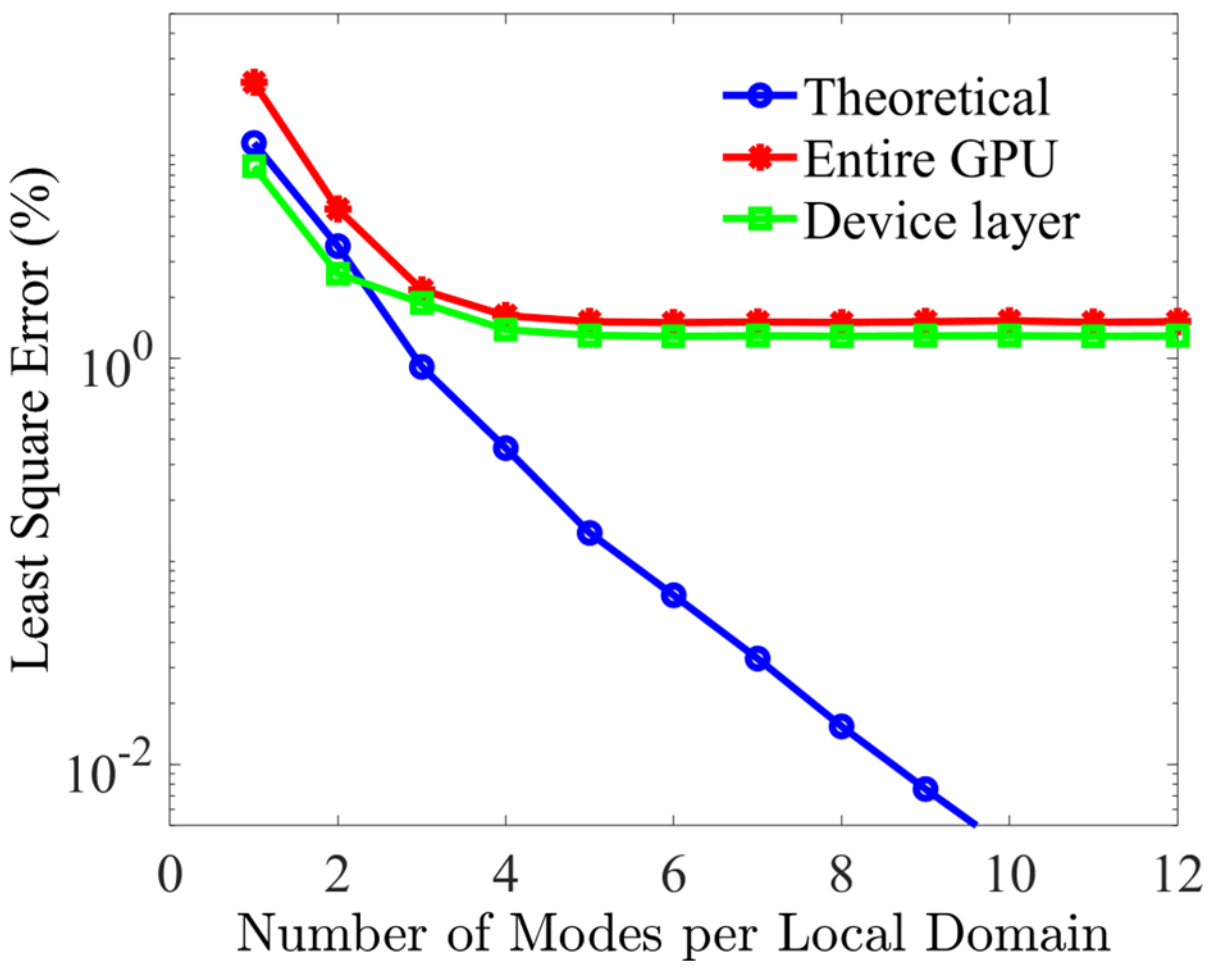

Fig. 7. *LSE* of thermal simulation of the Volta GV100 using LEnPOD-GP.

For LEnPOD-GP, DNSs of the generic local domain for each HSB is performed to train its IPOD-GP using random dynamic power excitations applied to the HSB. As shown in Fig. 4(a), the training data are utilized to solve POD modes and eigenvalues that are then applied to evaluate model parameters (i.e., matrix and vector elements in (4)) for each IPOD-GP. Based on the above consideration of thermal length and generic building blocks to select the generic truncated local domain (including the local domains whose HSBs are close to the processor edges), only 16 IPOD-GPs (i.e., $N_{gb}$ = 16) are needed to represent thermal solutions induced by 404 HSBs (i.e., 404 sets of ODEs, or $N_{gb}$ = 404) to construct LEnPOD-GP for the Tesla Volta GV100 GPU. Equivalent eigenvalues $\lambda_i^{eq}$ of LEnPOD-GP accounting for 404 sets of POD modes (404 truncated local domains) in this entire GPU evaluated from (6) represented by 16 I-PODGPs are shown Fig. 6. This eigenvalue spectrum declines rapidly to an extremely small value and becomes flattened beyond the 21th mode due to computer prevision. $LSE_{theo}$ estimated from (7) predicts an idea *LSE* near 3.2%, 0.9% and 0.36% with 2, 3 and 4 modes per truncated local domain, respectively, as shown in Fig. 7.

### *C. Demonstration of LEnPOD-GP for Tesla V100 Volta GV 100 GPU*

In the demonstration of LEnPOD-GP, the PM of the Tesla Volta™ GV100 GPU is adopted from [30], where a configurable GPU power simulator, AccelWattch, is developed and validated by GPU benchmarks. However, AccelWattch only generates the total power for each category of power components without the spatial power density distribution. To generate a PM for the demonstration used in Fig. 4(b), the total power dissipation of each category is randomly distributed among all components within the category based on the location of each component given in the floorplan. For example, the total power consumed by texture units obtained from AccelWattch is partitioned into 80 portions randomly, and each portion is assigned to all 4 texture units in each of 80 SMs (since all 4 texture units in each SM is taken as one HSB).

In this demonstration, 404 sets of ODEs given in (4) are first solved, as shown in Fig. 4(b), where each set includes *M* POD modes (i.e., *M* ODEs). Post processing is then performed in (5) using $\vec{a}_n$ to first obtain the dynamic temperature profile in each truncated domain $T_n(\vec{r},t)$ and then for the entire chip $T(\vec{r},t)$, as detailed in Fig. 4(b). The numerical *LSE* with respect to the FEniCS-FEM solution is presented in Fig. 7, compared with $LSE_{theo}$. Because the temperature responding to each HSB outside its truncated local domain is ignored for all 404 power source blocks and thermal gradients are relatively large in this case, *LSE* from LEnPOD-GP for the entire GPU is larger than $LSE_{theo}$, unlike EnPOD-GP for the quad-core CPU where *LSE* agrees well with $LSE_{theo}$ below 8 modes. Nevertheless, using a small number of modes per local domain, LEnPOD-GP still offers an accurate prediction of the thermal profile in the entire GPU with high thermal gradients and many crucial hot spots. For the entire domain, *LSE* in this case for the GPU shown in Fig. 7 reaches 2.18% or 1.6% with 3 or 4 modes per local domain, respectively, and remains at 1.5% beyond 4 modes. In most regions below the device (or heating) layer, temperature is low and close to the ambient, where the error tends to be larger. The LSE in the heating (device) layer, as shown in Fig. 7, is reduced to 1.87%, 1.39% or 1.3% when using 3, 4 or 5 modes, respectively. Dynamic temperature evolution at the intersection of Paths A and B (see Fig. 5) is given in Fig. 8(a). The temperature profiles along Paths A and B at $t$ = 25 ms are illustrated in Figs. 8(b) and 8(c), respectively. Using 3 or more modes, results derived from LEnPOD-GP agree quite well with those obtained from rigorous FEniCS-FEM.

The computational speedup (estimated in Intel Xeon Gold 6130 dual CPUs) for predicting the temperature in the entire GPU using LEnPOD-GP with 3 modes is around 900 times, compared to FEniCS-FEM. For the device layer, the efficiency improvement over FEniCS-FEM becomes 4,380 times. For applications related to thermal issues at the chip level of microprocessors, thermal information is usually only of interest in high temperature regions, i.e., in the high-power density cores and FUs. As mentioned above, differently from DNSs or thermal circuits, once the ODEs are solved (which is very fast) for LEnPOD-GP, one can select just a certain points in time or space to evaluate temperature from (5). When using 3 modes per HSB in LEnPOD-GP to calculate only the maximum temperature in the entire chip, a computational speedup of more than 1.1 or 0.7 million times over FEniCS-FEM can be achieved when 3 or 4 modes per HSB is implemented in LEnPOD-GP. The maximum error of the maximum chip temperature at all time steps predicted by LEnPOD-GP is near 1.2 ºC or 1.18 ºC when using 3 or 4 modes per HSB, respectively. These results indicate that, although the application of the truncated local domains in LEnPOD-GP slightly degrades the *LSE* in the entire processor, the maximum peak temperature in the entire chip remains accurate with a superior computational speed over the DNS.

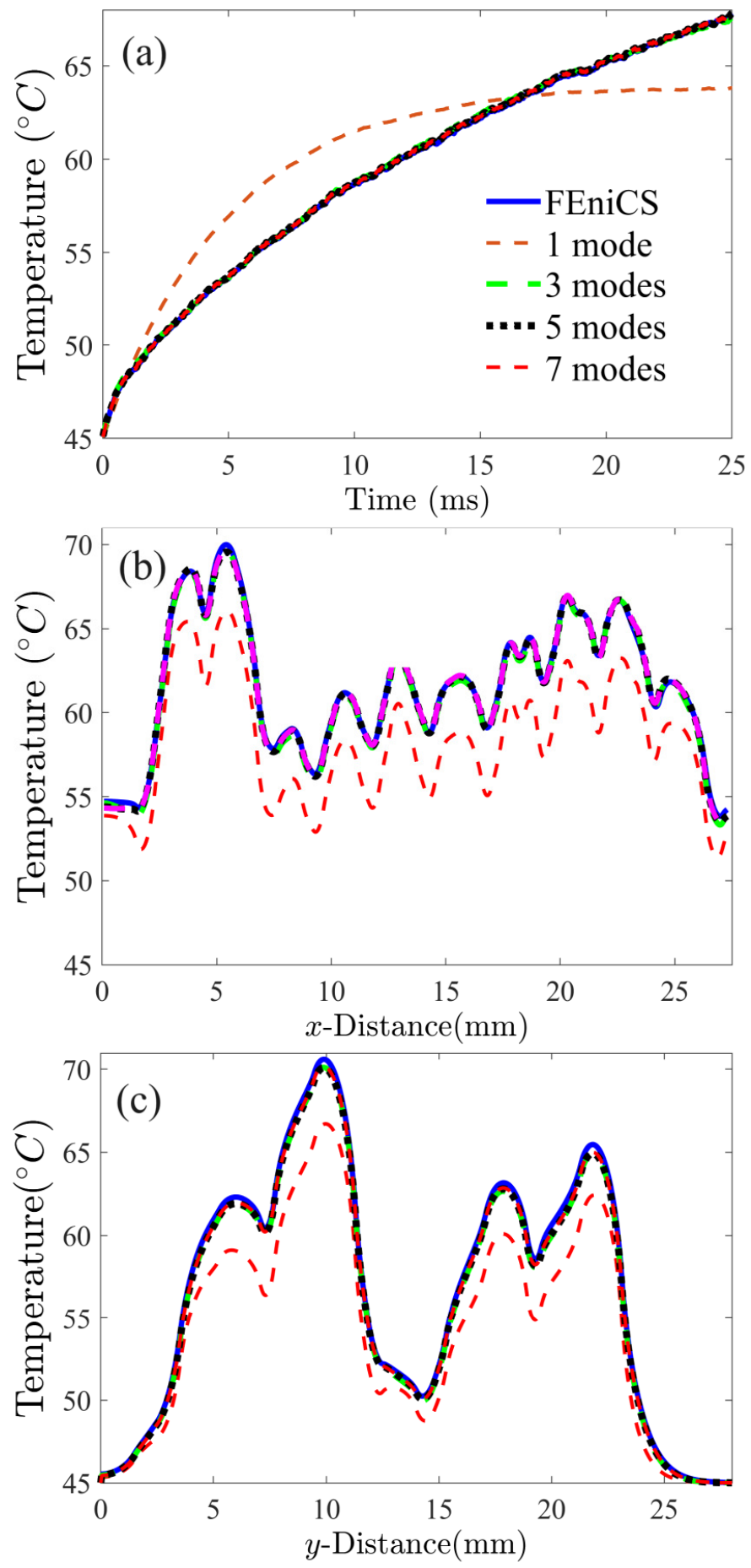

Fig. 8. (a) Dynamic temperature at the intersection of Paths A and B indicated in Fig. 5. (b) and (c) Temperature profiles at $t$ =25 ms along Path A and Path B, respectively, derived from FEniCS-FEM and LEnPOD-GP.

## V. Conclusions

The classical POD-GP simulation methodology suffers from intensive computational training of the POD modes in order to improve accuracy for microprocessors with a large number of cores. EnPOD-GP and LEnPOD-GP have been proposed to minimize the intensive training effort and improve simulation accuracy. EnPOD-GP has been applied to an AMD quad-core CPU with 9 HSBs each represented by a set of POD modes. It has been demonstrated in this case that EnPOD-GP is very accurate, efficient and robust to any dynamic PMs. Compared to FEniCS-FEM, an LSE near 0.78% is achieved with a 2,600-time computational speedup using 3 modes per HSB. To further minimize the training effort for processors with an enormous number of cores, LEnPOD-GP is developed, which applies local domain truncation for each HSB, together with generic building blocks, to reduce the massive amount of training data. In the demonstration of LEnPOD-GP on thermal simulation of Tesla Volta™ GV100 (a GPU with more than 13,000 cores), 16 generic truncated domains are trained to represent 404 truncated local domains for 404 HSBs. Even though the accuracy is degraded by neglecting temperature outside each of 404 truncated domains, the *LSE* is still as small as 1.87%, 1.39% or 1.3% in the device layer when using 3, 4 or 5 modes per HSB, respectively.

The saving in computational time to obtain the dynamic temperature distribution in the entire GPU using LEnPOD-GP with 3 modes per HSB for the selected GPU is near 900 times, compared to FEniCS-FEM. In the device layer, it is about 4,380 times. When evaluating only the peak temperature of the entire GPU at every time step, LEnPOD-GP offers a reduction in computational time over 1.1 million times, compared to FEniCS-FEM, with a maximum error near 1.2ºC. Since LEnPOD-GP does not need to post process dynamic temperature over the entire simulation space or time, the computational speedup will be even more for applications where the peak temperature is needed only at certain intervals of time.